\documentclass[10pt,aps,twocolumn,showpacs,pra]{revtex4-1}
\usepackage[dvips]{epsfig}
\usepackage{amssymb,amsfonts,amsmath,mathrsfs,mathbbol,color}
\usepackage{graphicx}

\begin{document}

\title{Optical control of resonant Auger processes}

\author{Antonio Pic\'{o}n}
\email[Corresponding author: ]{apicon@anl.gov}
\affiliation{Argonne National Laboratory, Argonne, Illinois 60439, USA}
\author{Christian Buth}
\affiliation{Argonne National Laboratory, Argonne, Illinois 60439, USA}
\author{Gilles Doumy}
\affiliation{Argonne National Laboratory, Argonne, Illinois 60439, USA}
\author{Bertold Kr\"assig}
\affiliation{Argonne National Laboratory, Argonne, Illinois 60439, USA}
\author{Linda Young}
\affiliation{Argonne National Laboratory, Argonne, Illinois 60439, USA}
\author{Stephen H. Southworth}
\affiliation{Argonne National Laboratory, Argonne, Illinois 60439, USA}
\date{\today}

\begin{abstract}
We theoretically show that core-excited state populations can be efficiently manipulated with strong optical fields during their decay, which takes place in a few femtoseconds. We focus on the $1s^{-1}3p$ resonant excitation in neon, where the $1s^{-1}3p$ and $1s^{-1}3s$ core-excited states are coupled by an optical field. By analyzing the Auger electron spectrum we observe the inner-shell population transfer induced by the optical coupling. We also show that the angular anisotropy of the Auger electron is imprinted in the multipeak structure induced by the optical-dressed continuum, namely sidebands.
\end{abstract}

\maketitle

\section{Introduction}

The understanding of inner-shell dynamics under strong optical fields \cite{Drescher2002,Meyer2012} is essential to develop alternative femtosecond x-ray characterization methods \cite{Dusterer2011} and schemes to control the x-ray interaction with atoms and molecules \cite{Buth2007,Glover2009}. In particular, the emergence of optical control schemes in the x-ray regime is of special interest to develop new approaches to tailor the x-ray pulse, reduce electronic damage in x-ray imaging experiments, and understand the optical-induced atom-specific dynamics. 

\begin{figure}[b] 
\centerline{\includegraphics[width=0.8\columnwidth,clip]{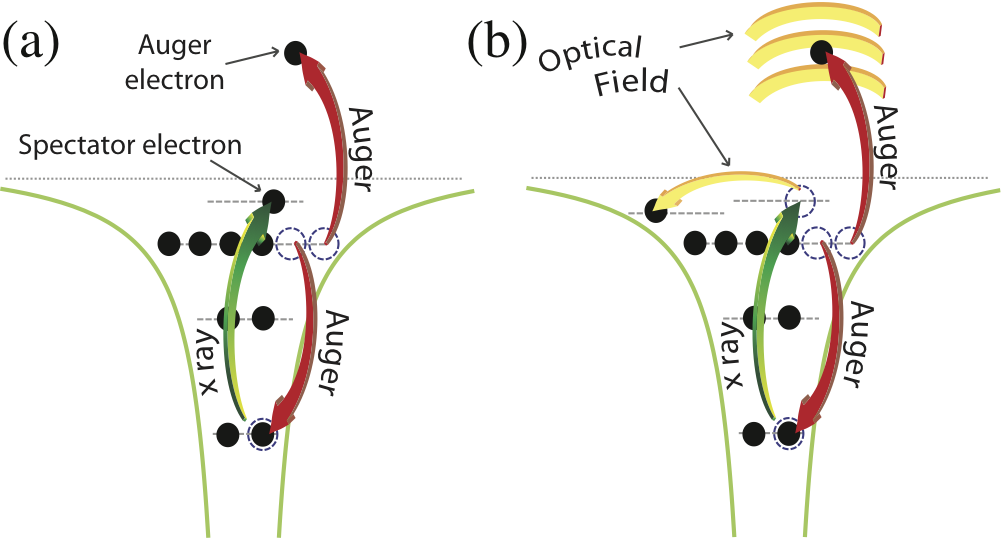}} 
\caption{(Color online)
(a) Resonant Auger process: A core electron is excited to an unoccupied Rydberg orbital generating a core-excited state with an inner-shell hole, which mainly decays by emitting an electron, namely Auger electron. (b) Resonant Auger process under an optical field: The optical laser field dresses the continuum and may couple resonantly two outer-shell Rydberg orbitals. 
}
\label{wavelength} \label{Fig_1}
\end{figure}

Resonant Auger processes have been extensively studied \cite{ReviewMehlhorn,ReviewArmen,ReviewPiancastelli} in the absence of optical fields. Here we address resonant Auger processes under strong optical fields, in which inner-shell dynamics is controlled by resonantly coupling two core-excited states (see Fig. \ref{Fig_1}). If the optical field is intense enough, the inner-shell excited-state populations can be manipulated in a few femtoseconds during their decay.
We focus on the resonant excitation $1s\!\rightarrow\!1s^{-1}3p$ in neon by x rays. Previous works \cite{Buth2007,Glover2009} demonstrated dramatic modification of the $1s\!\rightarrow\!1s^{-1}3p$ absorption cross section in the presence of a $\sim\!10^{13}$ W/cm$^2$ optical field due to the optical coupling between core-excited states. However, in the present work we show that such coupling induces observable effects in the Auger electron spectrum even with a $\sim\!10^{11}$ W/cm$^2$ optical field.

In this work, we identify two essential features in the resonant Auger electron spectrum in the presence of a strong optical field. First, the angular anisotropies are imprinted in the so-called (angle-integrated) sidebands, a multipeak structure in the Auger electron spectrum induced by the optical-dressed continuum \cite{Drescher2002,Meyer2012,Schins1994,Smirnova2003,Drescher2005,Kazansky2009,Buth2009,Kazansky2010}. Second, the optical field, besides dressing the continuum, also resonantly couples two core-excited states and efficiently transfers population among them. Hence the Auger electron spectrum is the result of the resonant Auger decay from both core-excited states. Furthermore, such femtosecond-scale optical manipulation permits the study of resonant Auger processes from core-excited states not accessible via one-photon transitions from the ground state.

{The work is organized as follows: In section \ref{sec:theoretical_model} we introduce our theoretical model to describe the resonant Auger process under an intense optical field. In section \ref{sec:results} we present the two main contributions of our work: the observation in the Auger electron spectrum of the optical manipulation of the core-excited states and of the resonant angular anisotropies in the sidebands. The conclusion of our paper is drawn in section \ref{sec:conlusion}.}

\section{Theoretical model} \label{sec:theoretical_model}

In our theoretical model for Ne, see appendices for a detailed description, the x rays couple the ground state ($\vert 0 \rangle$) with the Rydberg core-excited state $1s^{-1}3p$ ($\vert 1 \rangle$), while the optical field couples the Rydberg core-excited states $1s^{-1}3p$ and $1s^{-1}3s$ ($\vert 2 \rangle$). We apply a $\Lambda$-type three-level model, analogously to Ref. \cite{Buth2007}, to describe the coupling of the system with the x rays and the optical field, and the model is extended to account for the continuum states of the Auger electron, see appendix \ref{app:2}. The core-excited states may undergo resonant Auger decay into the states $\vert {\bf v}_a, i \rangle$, where $i$ represents the state of the final ion left after the decay with an Auger electron whose velocity is ${\bf v}_a$. The model also accounts for the resonant Auger features of linear dispersion and photon-bandwidth-dependent lineshape of the spectrum \cite{ReviewArmen,ReviewPiancastelli}. We derive the equations of motion (EOM) in the dipole and Markov approximations (or Wigner-Weisskopf theory) \cite{Wigner30,Knight90,Buth2009} (atomic units are used throughout)
\vspace{0.cm}
\begin{eqnarray}
&& \mathrm{i}\, {\dot a}_{0}(t) = [E_0  -\mathrm{i}{\Gamma_0\over2}] a_0(t) + \mu_{01} \varepsilon_{x} (t) a_1(t) \; , \nonumber \\
&& \mathrm{i}\, {\dot a}_{1}(t) = [E_1 + {\Delta_1\over2} -\mathrm{i}{\Gamma_1\over2}] a_1(t) + \nonumber \\
&& \hspace{1cm} \mu_{10} \varepsilon_{x} (t) a_0(t) + \mu_{12} \varepsilon_L (t) a_2(t) \; , \nonumber \\
&& \mathrm{i}\, {\dot a}_{2}(t) = [E_2 + {\Delta_2\over2} -\mathrm{i}{\Gamma_2\over2}] a_2(t) + \mu_{21} \varepsilon_L (t) a_1(t) \; , \nonumber \\
&& \mathrm{i}\, \dot{b}_i ({\bf v}_a,t) =  E_i({\bf v}_a) b_i({\bf v}_a, t) + \gamma_{i,1}({\bf v}_a) a_1 (t)  \nonumber \\ 
&& + \, \gamma_{i,2}({\bf v}_a) a_2 (t) +\! \int \!\! d{\bf v}_a'^{3} \; \mu_{i,i}({\bf v}_a;{\bf v}'_a) \varepsilon_L (t) b_i ({\bf v}'_{a}, t) \; ,  \label{EOM}
\end{eqnarray}
{where $a_0(t)$, $a_1(t)$, $a_2(t)$, and $b_i({\bf v}_a, t)$ represent the amplitudes of the ground, the Rydberg core-excited $1s^{-1}3p$, the Rydberg core-excited $1s^{-1}3s$, and the final ionic state with the Auger electron respectively, see more details in appendices \ref{app:2} and \ref{app:3}}. We assume Gaussian pulses for the x rays and a continuous wave for the optical field, see appendix \ref{app:1}. The energies of the ground and core-excited states are given by $E_{k}$, and for continuum states by $E_i({\bf v}_a)={\bf v}_a^2/2 + E_i^{+}$, where ${\bf v}_a^2/2$ is the Auger electron kinetic energy and $E_i^{+}$ is the final ion energy \cite{Coreno99,Schroeter99,Varma08}, see appendix \ref{app:3}. The dipole moment from state $a$ to $b$ is $\mu_{ba}$. The total decay $\Gamma_k$ accounts for the decay of the core-excited states due to non-radiative (e.g. Auger) and radiative processes as well as of the ionization induced by the optical laser and the x rays \cite{Aksela89,Hayaishi95,SchmidtBook,Coreno99}, see appendix \ref{app:2} for more details. The Stark shift of the core-excited states is given by the detuning $\Delta_k$. The optical ionization of the ground state is small due to the high ionization potential compared with the optical photon energy. The transition matrix elements $\gamma_{i,k}({\bf v}_a) \!=\! \langle {\bf v}_a,i\vert \hat{V}_{ee} \vert k \rangle$, $\hat{V}_{ee}$ being the electron-electron Coulomb interaction \cite{Buth2009}, do not depend on time, and within the Wigner-Weisskopf theory we can derive \cite{Buth2009}
\begin{eqnarray} \label{Partial_rate}
{\Gamma_i^{(k)}\over 2\pi} = 2\pi  \vert {\bf \tilde{v}}_a\vert \int_{-1}^{1} d(\cos\theta) \vert \gamma_{i,k}({\bf \tilde{v}}_a)\vert^2 \; ,
\end{eqnarray}
where $\Gamma_i^{(k)}$ is the partial rate of the core-excited state $k$ decaying into the final ion $i$, ${\bf \tilde{v}}_a$ is the velocity satisfying the energy conservation ${\bf \tilde{v}}_a^2/2 + E_i^{+} = E_k$, where $\theta$ is the angle between the x-ray polarization axis ${\bf e}_x$ and the velocity direction ${\bf \tilde{v}}_a$. The sum over all the possible decay channels is equal to the natural linewidth of the corresponding state, i.e. $\Gamma_k \!=\! \sum_i \Gamma_i^{(k)}$. Eq. (\ref{Partial_rate}) provides a relation between partial widths $\Gamma_i^{(k)}$ and $\gamma_{i,k}({\bf \tilde{v}}_a)$. However, for high kinetic energy Auger electrons, $\vert\gamma_{i,k}({\bf v}_a)\vert^2$ does not strongly depend on ${\bf v}_a$ \cite{ReviewArmen} and we can express $\gamma_{i,k}({\bf v}_a)$ in terms of partial widths $\Gamma_i^{(k)}$ as well. For linearly polarized light, within the dipole approximation, the differential partial rate of outgoing electrons has a simple form depending on an angular anisotropy parameter \cite{Yang48,Cooper89}, and from Eq. (\ref{Partial_rate}) we thus derive
\begin{eqnarray} \label{Auger_moment}
\gamma_{i,k}({\bf v}_a) = {e^{i\xi_i^{(k)}} \over\sqrt{(4\pi \vert {\bf \tilde{v}}_a\vert)}} \sqrt{\Gamma_i^{(k)} \over 2\pi}  \sqrt{1+\beta_i^{(k)} P_2(\cos\theta)} \; ,
\end{eqnarray}
up to a phase $\xi_i^{(k)}$, where $\beta_i^{(k)}$ is the angular parameter and $P_2(\cos\theta)$ is the second Legendre polynomial. Equation (\ref{Auger_moment}) satisfies Eq. (\ref{Partial_rate}) imposed by Wigner-Weisskopf theory. Analogously to $\Gamma_i^{(k)}$, $\beta_i^{(k)}$ can be obtained from experiments rather than resorting to numerical calculations. 

\begin{figure}[t] 
\centerline{\includegraphics[width=0.95\columnwidth,clip]{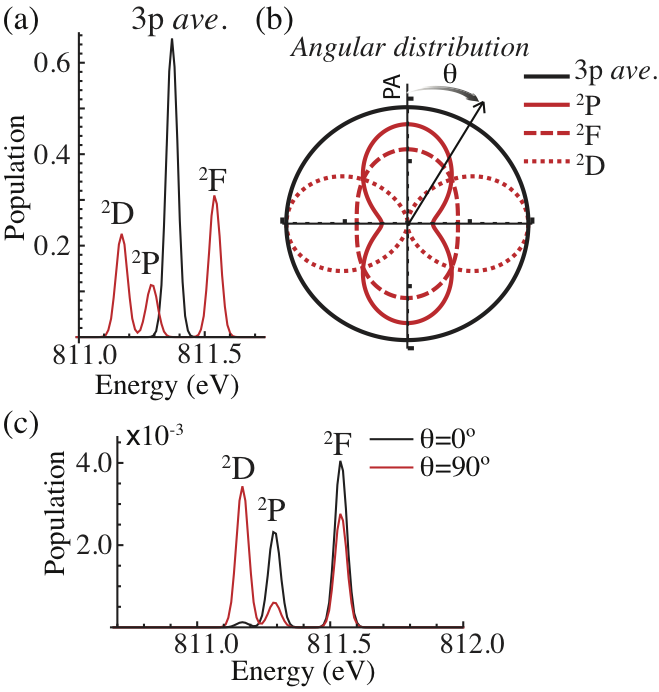}} 
\caption{(Color online)
(a) Angle-integrated Auger electron spectrum for the final ion state $2p^{-2}(^1D)3p$ with the three multiplets $^2 P$, $^2 D$, and $^2 F$. The notation 3p {\em ave}. stands for the average of the three multiplets (see more details in appendix \ref{app:3}). (b) Angular distributions of the multiplets with respect to the polarization axis of the x rays (PA). (c) Auger electron spectrum at angles (black line) $\theta=0^{\circ}$ and (red line) $\theta=90^{\circ}$.
}
\label{wavelength} \label{Fig_2}
\end{figure}

The core-excited state $1s^{-1}3p$ has a large number of Auger decay channels \cite{Aksela89,Hayaishi95}, and we only include in our model the main channel when the final ion state is $2p^{-2}(^1D)3p$, with the three multiplet splittings $^2 P$, $^2 D$, and $^2 F$ \cite{Aksela01,Shimizu2000}. The partial widths and the angular parameters for the multiplet are obtained from experiments \cite{Aksela89,Hayaishi95,Aksela01,Shimizu2000}, see appendix \ref{app:3}. For the core-excited state $1s^{-1}3s$, analogously to the $1s^{-1}3p$ case, the main channel is $2p^{-2}(^1D)3s$ (the spectator electron in the $3s$-orbital yields no energy splitting, we have only the $^2D$ term) and the partial width can be approximately obtained from the $1s^{-1}3p$ decay, see appendix \ref{app:3}. Since the $1s^{-1}3s$ state has total angular momentum $J=0$, the Auger angular distribution is then isotropic \cite{Cleff74,BookKabachnik}. Using linearly polarized weak x rays (resonant with the ground and the $1s^{-1}3p$ state transition), with no optical field, we obtain the angle-integrated Auger electron spectrum shown in Fig. \ref{Fig_2}(a) \cite{Aksela01}. In Fig. \ref{Fig_2}(b) we show the angular distributions for the terms $^2 P$, $^2 D$, and $^2 F$ with respect to the polarization axis (PA) of the x rays, and their Auger electron spectra at the angles $\theta=0^{\circ}$ and $\theta=90^{\circ}$ in Fig. \ref{Fig_2}(c) \cite{Shimizu2000}. The peaks in Fig. \ref{Fig_2} have linewidths which are narrower than the natural linewidth of 0.27 eV, due to the use of a narrow x-ray bandwidth of 0.033 eV \cite{ReviewPiancastelli}.

\begin{figure}[t] 
\centerline{\includegraphics[width=0.95\columnwidth,clip]{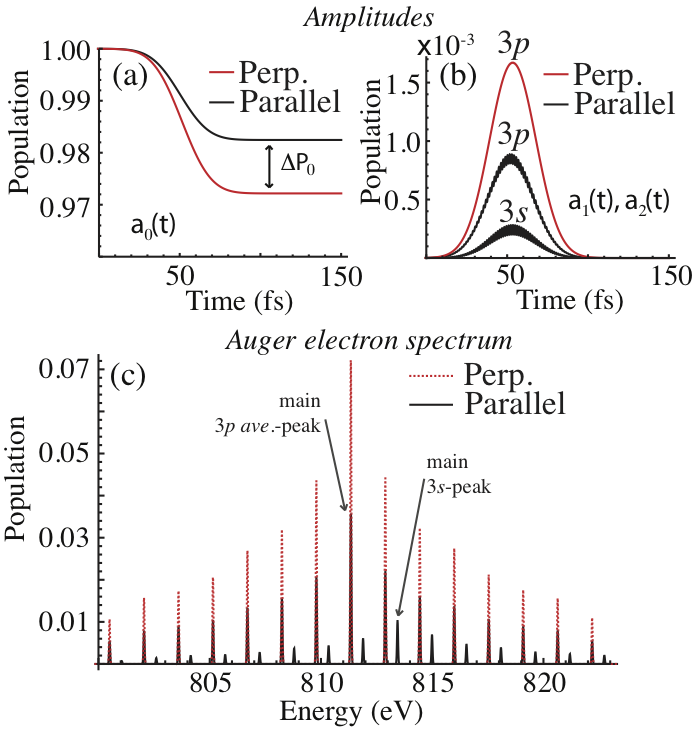}} 
\caption{(Color online)
(a) Evolution of the ground state population ($\vert a_{0}(t)\vert^2$). (b) Evolution of the population of the core-excited states $1s^{-1}3p$ and $1s^{-1}3s$ ($\vert a_{1}(t)\vert^2$ and $\vert a_{2}(t)\vert^2$). (c) Angle-integrated Auger electron spectrum in the presence of an optical field, see text for more details.
}
\label{wavelength} \label{Fig_3}
\end{figure}

\section{Results and Discussion} \label{sec:results}

{In this section we analyze the resonant Auger process under an intense optical field}. Figs. \ref{Fig_2}(a) and \ref{Fig_2}(b) show the spectrum and the angular distribution for the multiplet-averaged peak (denoted $3p\;ave.$), see appendix \ref{app:3}. Note that its angular distribution is practically isotropic. Now, with only two decay channels ($i\!=\!3p\;ave.$ and $i\!=\!3s$), we include a linearly-polarized 800 nm optical field (resonant with  $1s^{-1}3p \leftrightarrow 1s^{-1}3s$ within the natural linewidths and intensity $5\times10^{11}$ W/cm$^2$) for two different configurations; the polarization of the optical field and the x rays are parallel or perpendicular. Both polarization configurations are central to understanding the importance of the resonant coupling between Rydberg core-excited states in the Auger electron spectrum. In Fig. \ref{Fig_3}(a) and Fig. \ref{Fig_3}(b) we show the evolution of the ground state population ($\vert a_{0}(t)\vert^2$) and the core-excited state populations ($\vert a_{1}(t)\vert^2$ and $\vert a_{2}(t)\vert^2$) respectively; initially all the population is in the ground state, but it is weakly depleted after the arrival of the x rays, pumping up some population to the $1s^{-1}3p$ state. Despite the rapid decay of the core-excited states (lifetime 2.4 fs), in Fig. \ref{Fig_3}(b) we clearly observe that the optical field efficiently transfers some population to the $1s^{-1}3s$ state in the parallel configuration. Previously, the x-ray absorption cross section near the $1s\!\rightarrow\!1s^{-1}3p$ excitation was studied \cite{Buth2007} in the presence of a $10^{13}$ W/cm$^2$ optical field. It was shown that, in the perpendicular case, the optical field increases the linewidth of the absorption peak due to ionization, but does not mediate a coupling between $1s^{-1}3p$ and $1s^{-1}3s$ states because of selection rules. However, in the parallel case such coupling is mediated and, besides the increased linewidth, it also gives rise to a splitting that reduces the absorption cross section at the resonant frequency. Here, with a much lower optical intensity, we can observe this effect in Fig. \ref{Fig_3}(a), where the ground state depletion, related to the x-ray absorption cross section, is larger in the perpendicular configuration by a population difference $\Delta P_0$. Although this optical effect is small in the x-ray absorption cross section for the considered intensity, we can easily discern the resonant coupling between core-excited states by measuring the Auger electron spectrum, as the intensity is enough to transfer population to the $1s^{-1}3s$ state. We can calculate the angle-integrated Auger electron spectrum by performing the integral $p(E) = \int_0^\pi \!\! d\theta_i \int_0^{2\pi} \!\! d\varphi_i  \; \vert{\bf v}_a\vert \sin\theta_i \;  \vert b_i({\bf v}_a, \infty)\vert^2$, where $p(E)$ is the population at the energy $E={\bf v}_a^2 /2$ (see Fig. \ref{Fig_3}(c)). For the perpendicular configuration, the main $3p\;ave.$-peak is located at $\omega_X - E_{i=3p\;ave.}^{+}= 811.37$ eV, but additional peaks separated by an optical photon energy also appear. This multipeak structure, namely sidebands, is due to the continuum-continuum transitions, and its energy range is proportional to the square root of the optical intensity and the kinetic energy of the electron \cite{Schins1994}. In our case, the energy range spans more than 20 eV for $5\times10^{11}$ W/cm$^2$ due to the high kinetic energy of the Auger electron. However, for the parallel configuration where some population is transferred to the $1s^{-1}3s$ state, the Auger electron spectrum is the combination of the Auger decay from the $1s^{-1}3p$ and $1s^{-1}3s$ states dressed by the optical field, and it thus includes the main $3s$-peak at $\omega_X - \omega_L - E_{i=3s}^{+}= 813.47$ eV and its sidebands in Fig. \ref{Fig_3}(c).

\begin{figure}[t] 
\centerline{\includegraphics[width=0.95\columnwidth,clip]{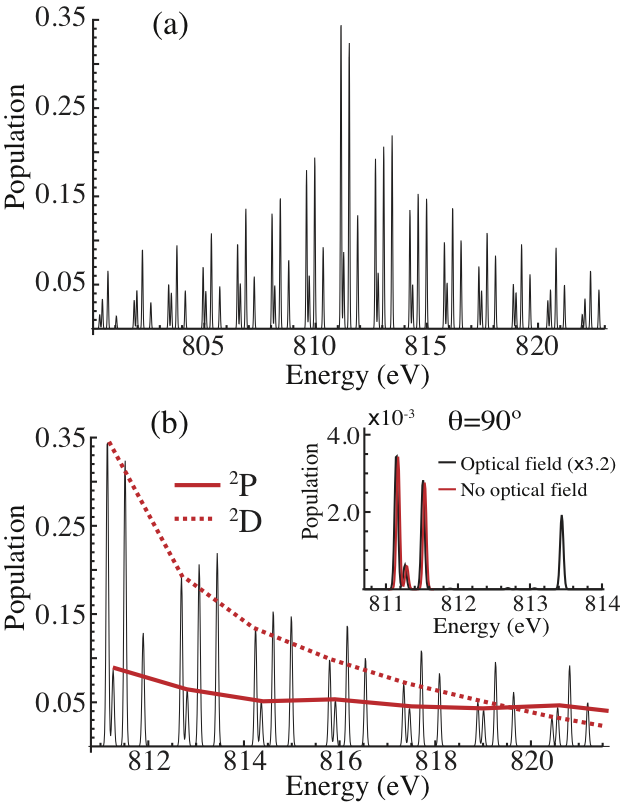}} 
\caption{(Color online)
(a) Angle-integrated Auger electron spectrum for $i=P,D,F$ and $3s$ channels. (b) As (a) but for a smaller energy range showing the envelope for the sidebands corresponding to the $^2P$ (solid line) and $^2D$ (dash line) terms. The inset shows the Auger spectrum measured at $\theta=90^{\circ}$, with (black line) and without (red line) optical field.
}
\label{wavelength} \label{Fig_4}
\end{figure}

We now consider the effects of the pronounced anisotropies in the multiplet terms $^2 P$, $^2 D$, and $^2 F$ of $2p^{-2}(^1D)3p$ (see Fig. \ref{Fig_2}). Sidebands have been theoretically and experimentally studied for normal Auger processes \cite{Schins1994,Drescher2002,Smirnova2003,Drescher2005,Kazansky2009,Kazansky2010,Buth2009,Meyer2012}. Nevertheless, the effects of the Auger angular distribution anisotropies on the structure of the sidebands have not been addressed. Sidebands strongly depend on the emission angle of the Auger electron with respect to the optical polarization axis (OPA). We show here that the angular parameter $\beta_{i}$ is manifested in the angle-integrated Auger electron spectrum, specifically in the envelope of sidebands. We emphasize that this effect is caused by the combination of the anisotropy of the Auger electron emission and the intrinsic anisotropy of the sidebands \cite{Meyer2012}. In Fig. \ref{Fig_3}(c) the channels $i\!=\!3p\;ave.$ and $i\!=\!3s$ had similar angular distributions and consequently similar sideband envelopes. Now, considering the multiplet, we expect a stronger optical dressing on the Auger electron from channel $i\!=\!P$ (angular distribution almost parallel to the OPA) rather than from channel $i\!=\!D$ (angular distribution almost perpendicular to the OPA). Figs. \ref{Fig_4}(a) and \ref{Fig_4}(b) show the angle-integrated Auger electron spectrum within 800-823 eV and 812-820 eV respectively. In Fig. \ref{Fig_4}(b) we observe the clear difference between the envelopes of the $D$-sidebands and the $P$-sidebands. The intensity of the $D$-sidebands decreases by a factor of ten in an energy range of 10 eV, while the $P$-sidebands only decrease by a factor of two in the same range. Furthermore, the effect of the anisotropies is also evident in comparing the relative heights of the three multiplet peaks, which are different at every optical photon energy. Finally we remark that the Auger electron spectrum detected perpendicular to the OPA does not present sidebands (see inset of Fig. \ref{Fig_4}(b)), where the same peaks of Fig. \ref{Fig_2}(c) (besides a small energy shift corresponding to the ponderomotive potential $U_p\!=\!0.03$ eV, analogous to the ATI phenomenon \cite{Prototapas1997}) are visible with an additional peak approximately at $813.47$ eV from the $1s^{-1}3s$ decay. This feature is of special interest because measurements perpendicular to the OPA can obtain the resonant Auger spectrum from the $1s^{-1}3s$ state without being obscured by sidebands.

\section{Conclusion} \label{sec:conlusion}

In conclusion, we have analyzed resonant Auger processes under a strong optical field. We focus on the $1s\!\rightarrow\!1s^{-1}3p$ excitation in Ne, accounting for a resonant coupling between two Rydberg core-excited states, $1s^{-1}3p$ and $1s^{-1}3s$, induced by the optical field. With no coupling between core-excited states, the $1s^{-1}3p$ decays into a continuum dressed by the optical field, yielding the multipeak structure of sidebands. The angular anisotropy parameter, defining the angular distribution of the Auger electron, is manifested in the envelope of the (angle-integrated) sidebands. With coupling between core-excited states, one can manipulate their population on the timescale of the natural lifetime if the intensity is large enough. Therefore the electron spectrum is the result of the resonant Auger decay from the core-excited $1s^{-1}3p$ and $1s^{-1}3s$ states dressed by the optical field. {In future investigations it would be interesting to consider intense x rays that induce Rabi oscillations between the ground and core-excited states, thus additionally modifying the Auger electron spectrum \cite{Nina2008,Kanter2011}.}





\begin{acknowledgments}
This work was supported by the Chemical Sciences, Geosciences, and Biosciences Division, Office of Basic Energy Sciences, Office of Science, U.S. Department of Energy, under Contract No. DE-AC02-06CH11357. A.P. acknowledges fruitful discussions with Shaohao Chen, Michal Tarana, and Elliot P. Kanter. 
\end{acknowledgments}

\appendix
\section{x rays and optical field} \label{app:1}
In all the numerical calculations, we use a Gaussian-envelope pulse for the x rays with electric field 
\begin{eqnarray}
\varepsilon_x (t) = \varepsilon_{0x} \exp[-(t\!-\!t_m)^2/2\sigma^2] \sin[\omega_X (t\!-\!t_m)] \; ,
\end{eqnarray}
where $\varepsilon_{0x}$ is the maximum amplitude of the electric field, $t_m$ is a given time when the electric field is maximum, $\sigma^2$ is the variance, and $\omega_X$ is the photon energy of the x rays. We have considered weak x rays with an intensity of $10^{14}$ W/cm$^2$. The variance is given by $\sigma=20$ fs, which defines a bandwidth of $0.033$ eV, smaller than the natural linewidth of the considered core-excited states. The photon energy is $\omega_X = 867.12$ eV, resonant with the ground and the $1s^{-1}3p$ state transition. {The narrow bandwidth is chosen to resolve the final ionic multiplet, see appendix \ref{app:3}. The multiplet structure was resolved experimentally, see for example Ref. \cite{Shimizu2000}}. 


For the optical field, we have considered a continuous wave with the electric field 
\begin{eqnarray}
\varepsilon_L (t)= \varepsilon_{0L} \sin[\omega_L(t-t_0)] \; ,
\end{eqnarray}
where $\varepsilon_{0L}$ is the maximum amplitude of the electric field, $t_0$ is a given time that defines the phase of the continuous wave, and $\omega_L$ is the photon energy of the optical field. We have considered a rather strong optical field with intensity $5\times10^{11}$ W/cm$^2$. The photon energy is chosen to be $\omega_L$=1.55 eV of a 800 nm Ti:sapphire laser.

Both x rays and optical field are linearly-polarized, using ${\bf e}_x$ for the x-ray polarization direction and ${\bf e}_L$ for the optical polarization direction.
{Both the pulse duration and the Gaussian shape of the x-ray pulse are not relevant in the considered physical scenario, due to the fact that we consider a CW optical laser and weak x rays, i.e. we are in the perturbative regime.}

In order to avoid numerical errors, at the beginning of the simulation the envelope of the optical field is zero and smoothly increases to be constant (continuous wave). Similarly, at the end the optical field envelope smoothly decreases to zero. Only when the optical field is an approximate continuous wave, the x-ray Gaussian pulse is sent to interact with the system. The Auger electron spectrum is calculated at the end of the interaction with the optical field.

\section{Ground state and core-excited states of neon} \label{app:2}

The main dynamics in neon is among the ground state ($\vert 0 \rangle$), the Rydberg core-excited state $1s^{-1}3p$ ($\vert 1 \rangle$), and the Rydberg core-excited state $1s^{-1}3s$ ($\vert 2 \rangle$), whose energies are obtained from experiments: $E_0=0$ eV, $E_1=867.12$ eV \cite{Coreno99}, and $E_2=865.34$ eV \cite{Schroeter99,Varma08}. The x rays couple the ground state with the $1s^{-1}3p$ state. The $1s^{-1}3p$ configuration has two states with $J=1$; $^1P_1$ and $^3P_1$. However the singlet state $^1P_1$ is the dominant excited state in Ne \cite{Shimizu2000}, when the total spin is preserved during the x-ray excitation. Therefore we only consider the singlet state for $\vert 1 \rangle$. The dipole moment is given by $\mu_{10} \!=\! \langle 1 \vert {\bf r}\cdot {\bf e}_x \vert 0 \rangle$, where ${\bf r}$ is the one-electron position operator and ${\bf e}_x$ is the polarization direction of the x rays. We obtain by means of {\em ab initio} calculations $\mu_{10}=0.0096$ a.u. \cite{FELLA}. The optical field couples the states $1s^{-1}3p$ and $1s^{-1}3s$. The $1s^{-1}3s$ configuration has two states; $^1S_0$ and $^3S_1$. As before, if the total spin of the system is preserved during the two-color excitation $1s\!\rightarrow\!1s^{-1}3s$, the dominant excited state is the singlet $^1S_0$. Therefore we only consider the singlet state for $\vert 2 \rangle$. The dipole moment is $\mu_{12} \!=\! \langle 1 \vert {\bf r}\cdot {\bf e}_L \vert 2 \rangle$, where here ${\bf e}_L$ is the polarization direction of the optical field. The {\em ab initio} calculations give $\mu_{12}=2.834$ a.u. \cite{FELLA}.

We consider the ionization and the Stark shift of the Rydberg core-excited states induced by the optical field via decay rates and detunings within the Markov approximation \cite{Knight90,Wigner30,Sakurai}, arisen by the optical coupling of states $\vert1\rangle$ and $\vert2\rangle$ with other states, mainly higher excited Rydberg and continuum states (the $1s^{-1}3p$ state is only $3.05$ eV from the continuum \cite{Coreno99}). We also account for the photoionization caused by the x rays via decay rates in first-order of perturbation theory. Hence, the decay rate $\Gamma_k$ of the $k$ state depends on the non-radiative and radiative decay processes $\Gamma_{1s}$ as well as of the ionization induced by the optical laser $\Gamma_k^{(ph)}$ and the x rays $\sigma_{k} J_{x}(t)$, i.e.
\begin{eqnarray}
\Gamma_k \!=\!\Gamma_{1s} (1-\delta_{0k})+ \Gamma_k^{(ph)} (t) + \sigma_{k} J_{x}(t) 
\end{eqnarray}
(for $k\!=\!0$ we do not have decay processes $\Gamma_{1s}$), where $\Gamma_{1s}$ is the natural linewidth of Ne$^+$ ($1s^{-1})$ (experimental value $\Gamma_{1s}\!=\!0.27$ eV, see references \cite{SchmidtBook,Coreno99}), $J_x(t)$ is the instantaneous x-ray flux \cite{McMorrowBook,ButhSantra07}, and $\sigma_{k}$ is the x-ray photoionization cross-section ($\sigma_0\!=\!2.3\times 10^{-20}$ cm$^2$ and $\sigma_1\!=\!\sigma_2\!=\!3.2\times 10^{-20}$ cm$^2$, obtained using \cite{LosAlamos1} and \cite{LosAlamos2}).

\begin{figure}[t] 
\centerline{\includegraphics[width=0.8\columnwidth,clip]{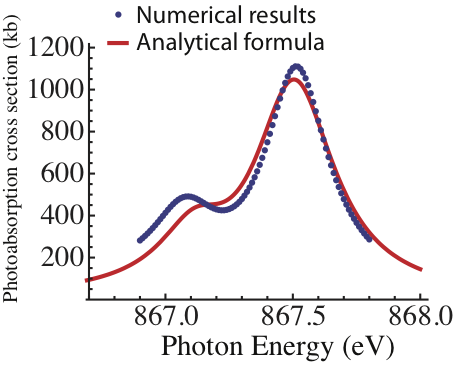}} 
\caption{(Color online)
X-ray photoabsorption cross section near the $1s\!\rightarrow\!1s^{-1}3p$ resonance for an 800 nm optical field with intensity $5\!\times\!10^{11}$ W/cm$^2$. Blue points, numerical results from the {\em ab initio} calculations \cite{FELLA}. Solid red line, results from the fitting using Eq. (2) of Ref. \cite{Buth2007}. 
}
\label{wavelength} \label{Fig_app}
\end{figure}

The optical photoionization $\Gamma_k^{(ph)}$ and Stark shift $S_k$ of the core-excited states are extracted by fitting {\em ab initio} calculations \cite{FELLA} for the x-ray absorption near the $1s\!\rightarrow\!1s^{-1}3p$ excitation with the formula given by Eq. (2) in Ref. \cite{Buth2007} (see also Ref. \cite{Harris1991}), where $\Gamma_k^{(ph)}$ and $S_k$ are assumed to be fitting parameters for a given intensity. In Fig. \ref{Fig_app} we show the fitting at the optical intensity of $5\times10^{11}$ W/cm$^2$ for the parameters $\Gamma_1^{(ph)}\!=\!0.03$ eV, $\Gamma_2^{(ph)}\!=\!0$, $S_1\!=\!0.14$ eV, and $S_2\!=\!0$. We note that this method is not highly precise, but it captures the main physics. The optical-induced Stark shift is small compared with the $1s\!\rightarrow\!1s^{-1}3p$ excitation energy, therefore the influence of this effect will slightly modify the x-ray absorption and consequently slightly modify the Auger electron yield. The optical ionization will promote the spectator electron to the continuum, hence the normal Auger decay occurs instead of the resonant one {\cite{Mazza2012}}. As the optical ionization rate is smaller than the corresponding resonant Auger decay rate, no significant contribution of normal Auger decay is expected in the calculated Auger electron spectrum. Finally, the optical ionization of the ground state is small due to the high ionization potential compared with the optical photon energy.

\section{Final ions after Auger decay in neon} \label{app:3}

The core-excited state may decay into a continuum state $\vert {\bf v}_a, i \rangle$, where $i$ represents the state of the final ion left after the decay with an Auger electron whose velocity is ${\bf v}_a$. Its energy is $E_i({\bf v}_a)={\bf v}_a^2/2 + E_i^{+}$, where ${\bf v}_a^2/2$ is the Auger electron kinetic energy and $E_i^{+}$ is the final ion energy. The strong-field approximation (SFA) is used, i.e. far from the nucleus the optical field interaction is dominant and Coulomb potential is not considered, and vice versa near the nucleus, where the optical laser is neglected (besides the optical-induced couplings mentioned above). The SFA, in which the optical field does not affect the transition matrix element $\gamma_{i,k}({\bf v}_a)\!=\! \langle {\bf v}_a,i\vert \hat{V}_{ee} \vert k \rangle$, is expected to be a good approximation for high kinetic energy electrons \cite{Kazansky2011}. We solve the differential equation for the continuum states $b_i ({\bf v}_a,t)$ (see Eq. (1) of the paper) similarly to Eq. (23) in Ref. \cite{Hergar2010}. This approach completely accounts for the continuum-continuum transitions responsible for sidebands.

The core-excited state $1s^{-1}3p$ has a large number of decay channels (i.e. different final ion states $\vert i\rangle$), with a significant contribution from shake-up processes \cite{Aksela01}. Here we only consider the main channel via Auger decay when the final ion state is $2p^{-2}(^1D)$ with a spectator electron in the $3p$-orbital. Hence, the final ion state $2p^{-2}(^1D)3p$ has a multiplet composed of three terms; $^2 P$, $^2 D$, and $^2 F$, whose energies are $E_{i=P}^{+}=55.83$ eV, $E_{i=D}^{+}=55.95$ eV, and $E_{i=F}^{+}=55.58$ eV, respectively \cite{Aksela01,Shimizu2000}. The experimental partial widths are given by $\Gamma_{i=P}^{(1)} = 0.015$ eV, $\Gamma_{i=D}^{(1)} = 0.030$ eV, and $\Gamma_{i=F}^{(1)} = 0.041$ eV \cite{Aksela89,Hayaishi95,Aksela01}, and the experimental angular anisotropy parameters by $\beta_{i=P}^{(1)} = 0.98$, $\beta_{i=D}^{(1)} = -0.95$, and $\beta_{i=F}^{(1)} = 0.27$ \cite{Shimizu2000}. Analogously to the $1s^{-1}3p$ case, the main channel of the core-excited $1s^{-1}3s$ state will be $2p^{-2}(^1D)3s$ (the $3s$ spectator electron yields no energy splitting, and we have only $^2D$ term) with a final energy $E_{i=3s}^{+}=52.10$ eV \cite{NIST}. Previously it was shown that the core-hole decay rate of $1s^{-1}3p$ is similar to that of $1s^{-1}$ \cite{Coreno99}, hence it is reasonable to assume that the core-hole decay rate of $1s^{-1}3p$ is similar to that of $1s^{-1}3s$. No experimental partial widths were found for the $1s^{-1}3s$ resonant Auger decay, however we assume $\Gamma_{i=3s}^{(2)}  \!=\! \Gamma_{i=P}^{(1)} +\Gamma_{i=D}^{(1)} +\Gamma_{i=F}^{(1)}  \!=\! 0.086$ eV because the $3s$ spectator electron does not introduce any multiplet splitting. In Ne, as the total spin of the system is preserved during the two-color excitation $1s\!\rightarrow\!1s^{-1}3s$, the decay state $1s^{-1}3s$ has total angular momentum $J=0$. Auger electrons from states with $J=0$ have an isotropic angular distribution \cite{Cleff74,BookKabachnik}, i.e. $\beta_{i=3s}^{(2)} = 0$. 

In our model, we also consider the average of the three $2p^{-2}(^1D)3p$ multiplets in order to reduce the number of peaks in the Auger electron spectrum and gain insight. We use the notation $i=3p\;ave.$ for this channel. We consider the average parameters of the three multiplets, i.e. the final ion energy $E_{i=3p\;ave.}^{+}\!\!= 55.75$ eV and the angular anisotropy parameter $\beta_{i=3p\;ave.}^{(1)}\!\!=-0.02$ (practically isotropic). The partial width simply is $\Gamma_{i=3p \,ave.}^{(1)}\!\!= \Gamma_{i=3s}^{(2)}\!=\!0.086$ eV.\\

We have not considered any effects of the optical field after Auger decay, such as Stark shift or ionization of the final ions, which are expected to be small. There are no resonant couplings between the considered final ion states because they are out of resonance with the optical photon energy. We note that these final ion states are metastable and they present narrow linewidths in contrast to core-excited states.

\end{document}